\documentclass{ifacconf}

\counterwithin*{section}{part}

\usepackage[english]{babel}
\usepackage{amssymb}
\usepackage{tikz}
\usepackage{mathtools}
\usepackage{etoolbox}

\usepackage[ruled,vlined,algo2e,linesnumbered]{algorithm2e}
\makeatletter
\patchcmd{\algocf@Vline}{\vrule}{\vrule\vspace{-.32em}}{}{}
\makeatother
\SetStartEndCondition{ }{ }{}%
\SetKw{KwTo}{to}\SetKwFor{For}{for}{\string do}{}%
\SetKwIF{If}{ElseIf}{Else}{if}{then}{elif}{else}{}%
\SetKwFor{While}{while}{\string do}{}%

\DeclareMathAlphabet{\pazocal}{OMS}{zplm}{m}{n}
\usepackage [autostyle, english = american]{csquotes} 
\MakeOuterQuote{"}

\usepackage{xargs}
\usepackage{soul, color, xcolor}
\usepackage[numbers]{natbib}

\usetikzlibrary{positioning,arrows,petri,calc,decorations.markings,arrows.meta}
\tikzset{
place/.style={circle,thick,minimum size=4mm,draw},
transitionV/.style={rectangle,thick,fill=black,minimum height=6mm,inner xsep=1pt}
}

\definecolor{myblue}{RGB}{0, 101, 202}
\definecolor{mygreen}{RGB}{130, 180, 0}
\definecolor{myred}{RGB}{197, 14, 31}
\definecolor{mypurple}{RGB}{128, 0, 128}
\definecolor{myyellow}{RGB}{204, 204, 0}
\definecolor{mygrey}{RGB}{105, 105, 105}

\newcommand{\N}{\mathbb{N}}
\newcommand{\No}{\mathbb{N}_0}
\newcommand{\Z}{\mathbb{Z}}

\newcommand{\R}{\mathbb{R}}
\newcommand{\Rmax}{{\R}_{\normalfont\fontsize{7pt}{11pt}\selectfont\mbox{max}}}
\newcommand{\Rmin}{{\R}_{\normalfont\fontsize{7pt}{11pt}\selectfont\mbox{min}}}
\newcommand{\Rbar}{\overline{\R}}

\newcommand{\floor}[1]{\left\lfloor#1\right\rfloor}

\renewcommand{\qed}{\hfill\ensuremath{\blacksquare}}

\newcommand{\places}{\pazocal{P}}
\newcommand{\transitions}{\pazocal{T}}
\newcommand{\arcs}{E}
\newcommand{\nodes}{N}

\makeatletter
\newcommand{\splus}{%
  \DOTSB\mathop{\mathpalette\mattos@splus\relax}\slimits@
}
\newcommand\mattos@splus[2]{%
  \vcenter{\hbox{%
    \sbox\z@{$#1\oplus$}%
    \resizebox{!}{0.9\dimexpr\ht\z@+\dp\z@}{\raisebox{\depth}{$\m@th#1\boxplus$}}%
  }}%
  \vphantom{\oplus}%
}
\makeatother

\makeatletter
\newcommand{\stimes}{%
  \DOTSB\mathop{\mathpalette\mattos@stimes\relax}\slimits@
}
\newcommand\mattos@stimes[2]{%
  \vcenter{\hbox{%
    \sbox\z@{$#1\otimes$}%
    \resizebox{!}{0.9\dimexpr\ht\z@+\dp\z@}{\raisebox{\depth}{$\m@th#1\boxtimes$}}%
  }}%
  \vphantom{\otimes}%
}
\makeatother

\makeatletter
\newcommand{\bigsplus}{%
  \DOTSB\mathop{\mathpalette\mattos@bigsplus\relax}\slimits@
}
\newcommand\mattos@bigsplus[2]{%
  \vcenter{\hbox{%
    \sbox\z@{$#1\sum$}%
    \resizebox{!}{0.9\dimexpr\ht\z@+\dp\z@}{\raisebox{\depth}{$\m@th#1\boxplus$}}%
  }}%
  \vphantom{\sum}%
}
\makeatother

\newcommand{\wA}{\mathsf{a}}
\newcommand{\wB}{\mathsf{b}}
\newcommand{\wC}{\mathsf{c}}
\newcommand{\wD}{\mathsf{d}}
\newcommand{\wZ}{\mathsf{z}}

\newcommand{\graph}{\pazocal{G}}
\newcommand{\nonegset}{\Gamma} 

\begin{document}
 

\begin{frontmatter}

\title{Weak Consistency of P-time Event Graphs\thanksref{footnoteinfo}} 

\author[1]{Davide Zorzenon}
\author[2]{Ji\v{r}\'{i} Balun}
\author[1,3]{J\"{o}rg Raisch}

\thanks[footnoteinfo]{Support from Deutsche Forschungsgemeinschaft (DFG) via grant RA 516/14-1 and under Germany’s Excellence Strategy -- EXC 2002/1 ``Science of Intelligence'' -- project number 390523135 is gratefully acknowledged.
Supported by the Ministry of Education, Youth and Sports under the INTER-EXCELLENCE project LTAUSA19098 and by IGA PrF 2022 018.
Supported by M\v{S}MT under the INTER-EXCELLENCE project LTAUSA19098 and by IGA PrF 2022 018.}

\address[1]{Control Systems Group, Technische Universit\"at Berlin, Germany (e-mail: [zorzenon,raisch]@control.tu-berlin.de)}
\address[2]{Faculty of Science, Palacky University in Olomouc, Czech Republic (e-mail: jiri.balun01@upol.cz)}
\address[3]{Science of Intelligence, Research Cluster of Excellence, Berlin, Germany}

\begin{abstract}
    P-time event graphs (P-TEGs) are event graphs where the residence time of tokens in places is bounded by specified time windows.
    In this paper, we define a new property of P-TEGs, called weak consistency.
    In weakly consistent P-TEGs, the amount of times a transition can fire before the first violation of a time constraint can be made as large as desired.
    We show the practical implications of this property and, based on previous results in graph theory, we formulate an algorithm of strongly polynomial time complexity that verifies it.
    From this algorithm, it is possible to determine, in pseudo-polynomial time, the maximum number of firings before the first constraint violation in a P-TEG.
\end{abstract}


\end{frontmatter}

\section{Introduction}

The class of event graphs in which places are associated with time intervals is called P-time event graphs (P-TEGs).
Whenever a token enters a place $p$ of a P-TEG, it must sojourn there for a time within the window associated to $p$; the dynamics of P-TEGs is thus nondeterministic.
Applications of P-TEGs arise in manufacturing, food industry, and chemical engineering, where sequence of tasks need to be completed in time in order to meet predefined specifications (see~\cite{becha2013modelling,spacek1999control,declerck2020critical}).

Like stability for linear systems, the most fundamental structural property of P-TEGs is arguably \textit{consistency}, i.e., the existence of an infinite sequence of firings of transitions that does not violate time constraints; such a sequence is then called a consistent trajectory.
If a P-TEG is not consistent, indeed, it is guaranteed that any trajectory will eventually violate one of the constraints.
Although this property has been investigated by several authors, no algorithm that provides necessary and sufficient conditions for consistency has yet been found.
In~\cite{Declerck2011FromExtremal}, the author analyzes consistency over finite event horizons $\{1,2,\ldots,K+1\}$, $K\in\No$, i.e., the existence of a sequence of $K+1$ firings that does not violate any time constraint.
When $K$ is finite and fixed, it is shown that consistency can be checked in polynomial time; additionally, sufficient conditions to efficiently verify consistency (over an infinite horizon) are given.
Other only sufficient and only necessary conditions for consistency have been presented in~\cite{vspavcek2021analysis}, where extremal periodic trajectories are considered.
In~\cite{zorzenon2020bounded}, another property, stronger than consistency, was defined: \textit{bounded consistency} (BC).
BC guarantees the existence of a consistent trajectory that does not lead to the accumulation of infinite delay between the $k$th firing of any pair of transitions, for all $k$.
An advantage of studying BC instead of consistency is that BC can be checked in strongly polynomial time (see~\cite{zorzenon2021periodic}).

However, even though BC is necessary for many applications (especially in \textit{continuous} production systems), for others this requirement could be too restrictive.
Take the example of \textit{intermittent} production systems, where the volume of production is limited and the mode of operations frequently changes.
If each mode of operation is modeled by a P-TEG, it may be not necessary to enforce each P-TEG to admit infinitely many firings, but only a finite -- although possibly arbitrary, i.e., not predefined -- number.
This prompts us to introduce a new property of P-TEGs, which characterizes this weaker requirement: \textit{weak consistency} (WC).
In a certain sense, WC extends the approach of~\cite{Declerck2011FromExtremal}; a weakly consistent P-TEG admits firing sequences over finite event horizons $\{1,\ldots,K+1\}$ where, however, $K$ can be taken as large as desired.
We show examples where this property does not imply consistency; indeed, in general the following implications hold:
\begin{equation}\label{eq:implications}
    \mbox{BC}\Rightarrow \mbox{consistency} \Rightarrow \mbox{WC}.
\end{equation}
Moreover, we prove that WC can be verified in strongly polynomial time, thanks to the reduction to a graph-theoretical problem studied in~\cite{hoeftig1995minimum}.
From~\eqref{eq:implications}, this allows, in some cases, to disprove consistency of P-TEGs in polynomial time.
Additionally, we formulate a pseudo-polynomial time algorithm that finds the largest finite event horizon before the first constraint violation (in the case a P-TEG is not weakly consistent); to the best of our knowledge, this is the first algorithm able to determine the length of the longest consistent trajectory in a P-TEG.

\subsection*{Notation}

The set of positive, respectively non-negative, integers is denoted by $\N$, respectively $\No$.
The set of non-negative real numbers is denoted by $\R_{\geq 0}$.
Moreover, $\Rmax \coloneqq \R \cup \{-\infty\}$, $\Rmin \coloneqq \R\cup\{\infty\}$, and $\Rbar \coloneqq \R\cup \{-\infty,\infty\}$.
If $A\in\Rbar^{n\times n}$, $A^\sharp$ indicates $-A^\intercal$, where $A^\intercal$ is the transpose of $A$.

\section{P-time event graphs}\label{se:2}

In this section, we recall the definition of P-time event graphs and their dynamics in the max-plus algebra.
After that, weak consistency is defined and compared to consistency and bounded consistency.

\subsection{General description and dynamics}

\begin{defn}[\cite{CALVEZ19971487}]
    An unweighted P-time Petri net is a 5-tuple $(\places,\transitions,\arcs,m,\iota)$, where $\places$ a finite set of places, $\transitions$ is a finite set of transitions, $\arcs\subseteq (\places\times \transitions)\cup (\transitions \times \places)$ is the set of arcs connecting places to transitions and transitions to places, and $m:\places\rightarrow\No$ and $\iota:\places\rightarrow\{[\tau^-,\tau^+]\ |\ \tau^-\in \R_{\geq 0},\tau^+\in\R_{\geq 0}\cup\{\infty\}\}$ are two maps that associate to each place $p\in\places$, respectively, its initial number of tokens (or marking) $m(p)$, and a time interval $\iota(p)=[\tau_p^-,\tau_p^+]$. 
\end{defn}

The dynamics of a P-time Petri net evolves as follows.
A transition $t\in\transitions$ is said to be enabled if either it has no upstream places (i.e., $(\nexists p\in \places)\ (p,t)\in\arcs$) or each upstream place $p\in \places$ contains at least one token that has resided in $p$ for a time included in interval $[\tau_p^-,\tau_p^+]$.
When transition $t$ is enabled, it can fire, causing one token to be instantaneously removed from each upstream place ($(p\in \places) \ (p,t)\in \arcs$) and one token to be instantaneously added to each downstream place ($(p\in \places) \ (t,p)\in\arcs$) of $t$.
If a token resides for too long in $p$, violating the constraint imposed by interval $[\tau_p^-,\tau_p^+]$, then the token is said to be dead.
As we will see later, in some P-time Petri nets this event cannot be avoided by carefully choosing the firing time of each transition; consequently, these P-time Petri nets will be called inconsistent.

In this paper, we focus on a subclass of P-time Petri nets called P-time event graphs (P-TEGs).
A P-TEG is a P-time Petri net where each place has exactly one upstream and one downstream transition (i.e., $(\forall p\in \places)\ \exists ! (t_{\textup{up}},t_{\textup{down}})\in\transitions\times \transitions:\ (t_{\textup{up}},p)\in\arcs\wedge (p,t_{\textup{down}})\in\arcs$).
We will consider only P-TEGs in which, for each place, the initial number of tokens is either 0 or 1; this will not affect the generality of the results, as in~\cite{amari2005control} it has been shown that every P-TEG can be transformed into an equivalent one that satisfies this property.
This assumption allows us to formulate the dynamics of a P-TEG with $|\transitions| = n$ transitions as follows.
Let $A^0,A^1\in \Rmax^{n\times n}$, $B^0,B^1\in \Rmin^{n\times n}$ be four matrices such that, if there exists a place $p$ with initial marking $\mu\in\{0,1\}$, upstream transition $t_j$ and downstream transition $t_i$, then $A^\mu_{ij} = \tau_p^-$ and $B^\mu_{ij}=\tau_p^+$, otherwise $A^\mu_{ij} = -\infty$ and $B^\mu_{ij} = \infty$.
The dater function $x:\No\rightarrow \R^n$ represents the firing time of transitions; element $x_i(k)$ is the time of the $(k+1)$st firing of transition $t_i$.
Since the $k$th firing of any transition $t_i\in\transitions$ cannot occur before the $(k+1)$st, it is natural to assume that $x_i(k-1) \leq x_i(k)$ for all $k\in\N$.
The evolution of a P-TEG can now be described by the following set of inequalities (for a more detailed explanation, see~\cite{vspavcek2021analysis}): $\forall k\in\No$, $i\in\{1,\ldots,n\}$,
\begin{equation}\label{eq:dynamics_PTEGs}
	\left\{
	\begin{array}{rcl}
		\max_{j = 1,\dots,n} A^0_{ij}+ x_j(k) \leq & x_i(k) & \leq \min_{j = 1,\dots,n} B^0_{ij} + x_j(k)\\
		\max_{j = 1,\dots,n} A^1_{ij} + x_j(k) \leq & x_i(k+1) & \leq \min_{j = 1,\ldots,n} B^1_{ij} + x_j(k)
	\end{array}
	\right.
	.
\end{equation}
In particular, an (infinite) trajectory $\{x(k)\}_{k\in\No}$ of the dater function is called consistent if it satisfies~\eqref{eq:dynamics_PTEGs} for all $k\in\No$, as it represents an evolution of the P-TEG for which no token dies.
We also say that the \textit{finite} trajectory $\{x(k)\}_{k\in \{0,1,\ldots,K\}}$ of length $K\in\No$ is consistent, if satisfies the first inequality of~\eqref{eq:dynamics_PTEGs} for all $k\in\{0,1,\ldots,K\}$, and the second inequality for all $k\in\{0,1,\ldots,K-1\}$.
When not otherwise stated, with the term "trajectory" we will always refer to an infinite trajectory of the dater function.

\subsection{Dynamics in the max-plus algebra}

Inequalities of the form~\eqref{eq:dynamics_PTEGs} can be conveniently studied in the max-plus algebra, where they take the name of \textit{linear-dual inequalities}.
Let $a,b\in\Rbar$.
We define the max-plus operations of addition $\oplus$, multiplication $\otimes$, dual addition $\splus$, and dual multiplication $\stimes$ as follows:
\[
    a\oplus b = \max(a,b),\quad a \otimes b = 
    \begin{dcases}
        a + b & \mbox{if } a,b \neq -\infty,\\
        -\infty & \mbox{otherwise},
    \end{dcases}
\]\[
    a\splus b = \min(a,b),\quad a\stimes b =     
    \begin{dcases}
        a + b & \mbox{if } a,b \neq \infty,\\
        \infty & \mbox{otherwise}.
    \end{dcases}
\]
The operations can be extended to matrices in the usual way; given $A,B\in\Rbar^{m\times n}$, $C\in\Rbar^{n\times p}$, for all $i\in\{1,\ldots,m\}$, $j\in\{1,\ldots,n\}$, $h\in\{1,\ldots,p\}$,
\[
    (A\oplus B)_{ij} = A_{ij} \oplus B_{ij},\quad (A\otimes C)_{ih} = \bigoplus_{k = 1}^n A_{ik} \otimes C_{kh},
\]
\[
    (A\splus B)_{ij} = A_{ij} \splus B_{ij},\quad (A\stimes C)_{ih} = \bigsplus_{k = 1}^n A_{ik} \stimes C_{kh}.
\]
The partial order relation $\preceq$ between two matrices of the same dimension is induced by $\oplus$ as: $A\preceq B \Leftrightarrow A \oplus B = B$; hence, $A\preceq B$ is equivalent to $A_{ij} \leq B_{ij}$ for all indices $i$, $j$.

We now have all the ingredients to rewrite~\eqref{eq:dynamics_PTEGs} in the max-plus algebra:
\begin{equation}\label{eq:dynamics_PTEGs_maxplus}
\forall k\in\No, \quad 
	\left\{
	\begin{array}{rcl}
		A^0\otimes x(k) \preceq & x(k) & \preceq B^0\stimes x(k)\\
		A^1\otimes x(k) \preceq & x(k+1) & \preceq B^1\stimes x(k)
	\end{array}
	\right.
	~.
\end{equation}
Rather than representing a mere cosmetic change, the reformulation of P-TEGs dynamics in the max-plus algebra has the advantage to make classical algebraic and graph-theoretical results accessible to the study of P-TEGs, as we will see in Section~\ref{se:4}.

\section{Structural properties}\label{se:3}

In this section, the definitions of consistency and bounded consistency are recalled from~\cite{zorzenon2020bounded}, and weak consistency is presented.
The three structural properties are then compared by means of simple examples of P-TEGs.

\begin{defn}[Consistency]
    A P-TEG is said to be consistent if it admits a consistent trajectory $\{x(k)\}_{k\in\No}$.
\end{defn}

\begin{defn}[Bounded consistency (BC)]
    A P-TEG is said to be boundedly consistent if it admits a consistent trajectory $\{x(k)\}_{k\in\No}$ such that $(\exists M\in\R_{\geq 0}) (\forall i,j\in \{1,\ldots,n\}, k\in\No) \ x_i(k) - x_j(k) \leq M$.
\end{defn}

Consider a production system modeled by a P-TEG, where the $k$th firing of a transition indicates the start or finish of a sub-task, and the $k$th final product is completed when all the transitions have fired for the $k$th time.
Note that the difference $x_i(k) - x_j(k)$ represents the delay of the $(k+1)$st firing of transition $t_i$ with respect to the $(k+1)$st firing of transition $t_j$.
BC narrows the field of interest to only those trajectories where this quantity is bounded, i.e., no infinite accumulation of delay is possible.
In continuous production systems, where large amounts of products are processed without interruption, these are usually the only acceptable trajectories.
The accumulation of delay between two sub-tasks $t_i$ and $t_j$ would otherwise cause an unbounded increase in the time required for producing the $k$th final product.
On the other hand, in intermittent production systems, characterized by a production based on customer's demand and more frequent shut downs, less stringent requirements may be acceptable.
A desirable property of the system could be, in this case, the capability to complete $K$ products without interruptions or constraint violations, for \textit{any} given value of $K$.
This would make it possible to schedule the production flow following the arrival of a new order, being certain that the system will be able to meet the customer's demand for any quantity of products required.
In mathematical terms, this requirement is formalized as follows.

\begin{defn}[Weak consistency (WC)]
    A P-TEG is said to be weakly consistent if $\forall K\in\No$ there exists a consistent finite trajectory $\{x(k)\}_{k\in\{0,1,\ldots,K\}}$. 
\end{defn}

We remark that the "$\forall$" in the definition distinguishes the property of WC from the one studied in~\cite{Declerck2011FromExtremal}, where the considered finite event horizons are fixed a priori.

Clearly, consistency implies WC.
At a superficial glance, it might even seem that WC coincides with consistency; after all, if a finite, but as-long-as-desired trajectory $x(0),x(1),\ldots,x(K)$ exists that does not violate any time-window constraint, what prohibits extending it indefinitely to $x(K+1),x(K+2),\ldots$?
As the examples presented in the next subsection will reveal, this is however not always possible.

\subsection{Some simple examples}\label{su:examples}

\begin{figure}
	\centering
	\resizebox{.5\linewidth}{!}{
		\begin{tikzpicture}[node distance=.5cm and 1.5cm,>=stealth',bend angle=30,thick]
\footnotesize
\node[transitionV,label=below:{$t_1$}] (t1) {};
\node[place,right=of t1,label=above:{$[0,\gamma_\wZ]$}] (p12) {};
\node[transitionV,right=of p12,label=below:{$t_2$}] (t2) {};
\node[place,tokens=1,above=of t1,label=above:{$[\alpha_\wZ,\alpha_\wZ]$}] (p11) {};
\node[place,tokens=1,above=of t2,label=above:{$[\beta_\wZ,\beta_\wZ]$}] (p22) {};

\draw (t1) edge[->] (p12);
\draw (p12) edge[->] (t2);
\draw (t1.90-15) edge[bend right,->] (p11);
\draw (p11) edge[bend right,->] (t1.90+15);
\draw (t2.90-15) edge[bend right,->] (p22);
\draw (p22) edge[bend right,->] (t2.90+15);

\end{tikzpicture}
	}
	\caption{Example of P-TEG $\pazocal{P}_\wZ$.}
	\label{fig:P-TEG_example}
\end{figure}
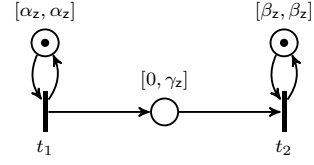
\begin{table}[t]
	\begin{center}
	\caption{Parameters for the P-TEG of Figure~\ref{fig:P-TEG_example}.} \label{tab:P-TEG_parameters}
		\begin{tabular}{cccc}
			$\wZ$ & $\alpha_\wZ$ & $\beta_\wZ$ & $\gamma_\wZ$ \\\hline
			$\wA$ & 1 & 1 & $\infty$ \\
			$\wB$ & 1 & 2 & $\infty$ \\ 
			$\wC$ & 2 & 1 & $\infty$ \\
			$\wD$ & 2 & 1 & 10
		\end{tabular}
	\end{center}
\end{table}

Consider the family of P-TEGs $\pazocal{P}_\wZ$ represented in Figure~\ref{fig:P-TEG_example}, where time intervals are parametrized with respect to label $\wZ$; the values of time windows are given for $\wZ\in\{\wA,\wB,\wC,\wD\}$ in Table~\ref{tab:P-TEG_parameters}.
The following analysis will show that the four examples represent all possible combinations of P-TEGs structural properties: $\pazocal{P}_\wA$ is boundedly consistent, $\pazocal{P}_\wB$ is consistent but not boundedly consistent, $\pazocal{P}_\wC$ is weakly consistent but not consistent, and $\pazocal{P}_\wD$ is not weakly consistent.

The matrices $A^0,A^1,B^0,B^1$ characterizing the P-TEG labeled $\wZ$ are:
\[
	A^0_\wZ = 
	\begin{bmatrix}
		-\infty & -\infty \\
		0 & -\infty
	\end{bmatrix},\quad
	A^1_\wZ = 
	\begin{bmatrix}
		\alpha_\wZ & -\infty \\
		-\infty & \beta_\wZ
	\end{bmatrix},
\]\[
	B^0_\wZ = 
	\begin{bmatrix}
		\infty & \infty \\
		\gamma_\wZ & \infty
	\end{bmatrix},\quad
	B^1_\wZ = 
	\begin{bmatrix}
		\alpha_\wZ & \infty \\
		\infty & \beta_\wZ
	\end{bmatrix}.
\]
The analysis of $\pazocal{P}_\wZ$ is made particularly simple by the fact that lower and upper bound constraints coincide in the two places with an initial token.
In contrast to the general case, this forces the dynamics of the P-TEGs to evolve deterministically, once $x(0)$ is set.
Indeed, from~\eqref{eq:dynamics_PTEGs}, the dater function must satisfy the following conditions (written in standard algebra) for all $k\in\No$:
\[
	\begin{bmatrix}
		x_1(k+1)\\
		x_2(k+1)
	\end{bmatrix}
	=
	\begin{bmatrix}
		x_1(k) + \alpha_\wZ\\
		x_2(k) + \beta_\wZ
	\end{bmatrix},\;
	x_1(k) \leq x_2(k) \leq x_1(k) + \gamma_\wZ.
\]
For $\pazocal{P}_\wA$, they imply $\{x(k)\}_{k\in\No} = \{x(0) + [k\ k]^\intercal\}_{k\in\No}$; clearly, a trajectory of this kind satisfies the definition of BC for all $M\geq x_2(0) - x_1(0) \in\R_{\geq 0}$.
The only consistent trajectories for $\pazocal{P}_\wB$ have the form $\{x(k)\}_{k\in\No} = \{x(0) + [k\ 2k]^\intercal\}_{k\in\No}$; thus, the delay $x_2(k) - x_1(k)$ is an increasing function of $k$ that grows beyond all bounds for $k\rightarrow \infty$, which implies that $\pazocal{P}_\wB$ is not boundedly consistent.
The situation is different in $\pazocal{P}_\wC$; taking $x_2(0) = x_1(0) + \kappa$, $\kappa \in\R_{\geq 0}$, trajectory $\{x(k)\}_{k\in\No} = \{x(0) + [2k\ k]^\intercal\}_{k\in\No}$ is consistent only for the first $K = \floor{\kappa}$ values of $k$.
Indeed, for $k = K + 1$, $x_1(k) > x_2(k)$, which violates the dynamical inequalities of the P-TEG causing the first token death.
However, since $\kappa$ can be chosen as large as desired, the number of valid transition firings before the first death occurs can be extended at will.
On the other hand, in $\pazocal{P}_\wD$ the value of $\gamma_\wD$ imposes $\kappa$ to be $\leq 10$; the consequence is that there can be at most $K+1 = 11$ valid firings of each transition. 
$\pazocal{P}_\wD$ is thus not weakly consistent.

Observe that, by choosing larger but finite values for $\gamma_\wD$, the length of the longest consistent finite trajectory for $\pazocal{P}_\wD$ increases.
This makes it reasonable to think that, if an algorithm that verifies WC exists, then its time complexity is pseudo polynomial in the best case, as it shall depend (hopefully in a polynomial way) on the magnitude of the elements of $A^0,A^1,B^0,B^1$; indeed, the algorithm should be able to distinguish instances of P-TEGs as $\pazocal{P}_\wC$ from those like $\pazocal{P}_\wD$, in which arbitrary large values of $\gamma_\wD$ can be considered.
Instead, quite surprisingly, a strongly polynomial time algorithm that checks WC exists; showing this will be the focus of the next section.

\section{Verification of weak consistency}\label{se:4}

We start this section by summarizing some facts that connect the max-plus algebra with graph theory; for more details on this topic, we refer to~\cite{baccelli1992synchronization,butkovivc2010max}.
These facts will be needed to introduce the strongly polynomial time algorithm that checks WC.

\subsection{Max-plus algebra and graph theory}

\begin{defn}[Precedence graph]
    Let $A\in\Rmax^{n\times n}$.
    The precedence graph corresponding to matrix $A$ is the pair $\graph(A) = (\nodes,\arcs)$, where $\nodes = \{1,\ldots,n\}$ is the set of nodes, and $\arcs \subseteq \nodes\times \nodes$ is the set of weighted arcs, defined such that there is an arc $(j,i)\in\arcs$ with weight $A_{ij}$ if and only if $A_{ij}\neq -\infty$.
\end{defn}

A path on $\graph(A)$ is a sequence of nodes $\rho = (i_1,i_2,\ldots,i_{r+1})$ such that $(i_j,i_{j+1})\in\arcs$ for all $j\in\{1,\ldots,r\}$; the number $|\rho|=r\in\N$ is the length of $\rho$.
The weight $w_\rho$ of path $\rho$ is the sum (in standard algebra) of the weights of the arcs composing it; in the max-plus algebra, this quantity can be computed as $\bigotimes_{j = 1}^{|\rho|} A_{i_{j+1}i_j}$.
The path $\rho$ is a circuit if $i_1 = i_{|\rho|+1}$.
There is a correspondence between powers of max-plus matrices and weight of paths in precedence graphs; given $r\in\N$, let $A^{\otimes r}$ be recursively defined by $A^{\otimes 0} = E_\otimes$ (where $(E_\otimes)_{ij} = 0$ if $i=j$, $(E_\otimes)_{ij} = -\infty$ if $i\neq j$), $A^{\otimes r} = A^{\otimes r-1}\otimes A$.
Then, element $(A^{\otimes r})_{ij}$ is equal to the largest weight of all paths from node $j$ to node $i$ of length $r$ in $\graph(A)$.
We denote by $\nonegset$ the set of all precedence graphs that do not contain circuits with positive weight; formally, $\nonegset = \{\graph(A)\ |\ (\exists n\in\N)\ A\in\Rmax^{n\times n} \mbox{ and } (\forall r \in \N, i\in\{1,\ldots,n\}) \ (A^{\otimes r})_{ii} \leq 0\}$.
Moreover, the Kleene star operator is defined by $A^* = \bigoplus_{r \in \No} A^{\otimes r}$; the result of this operation can be computed in $\pazocal{O}(n^3)$ whenever $\graph(A)\in\nonegset$, and the property "$\graph(A)\in\nonegset$" can be verified in $\pazocal{O}(n^3)$.

\begin{prop}[\cite{butkovivc2010max}]\label{pr:inequalities_and_graphs}
    Let $A\in\Rmax^{n\times n}$.
    Inequality $A\otimes x\preceq x$ admits a solution $x\in\R^n$ if and only if $\graph(A)\in\nonegset$.
    Moreover, in this case the solution set $\{x\in\R^n\ |\ A\otimes x\preceq x\}$ coincides with $\{A^*\otimes u\ |\ u\in\R^n\}$.
\end{prop}

In the next section, WC will be rephrased as a system of infinitely many inequalities of the form $A\otimes x\preceq x$.
To do this, we need the following preliminary result.

\begin{prop}[\cite{cuninghame2012minimax}]\label{pr:inequalities}
    Let $x,y\in\R^n$, $A,B\in \Rmax^{n\times n}$.
    Then,
    \[
        x \preceq A^\sharp \stimes y\quad \Leftrightarrow \quad     A\otimes x \preceq y,
    \]
    and
    \[
	\left\{
	\begin{array}{rl}
		A\otimes x \preceq & y \\
		B\otimes x \preceq & y 
	\end{array}
	\right.
        \quad \Leftrightarrow \quad (A\oplus B)\otimes x\preceq y.
    \]
\end{prop}

\subsection{The polynomial-time algorithm}

We start by rephrasing the conditions for WC in the max-plus algebra.
A P-TEG is weakly consistent if and only if, $\forall K\in \No$, the following set of inequalities admits solutions $x(0),\ldots,x(K)\in\R^n$: 
\begin{equation*}
	\begin{array}{rcll}
		A^0\otimes x(k) \preceq & x(k) & \preceq B^0\stimes x(k) & \,
\forall k\in\{0,\ldots,K\}, \\
		A^1\otimes x(k) \preceq & x(k+1) & \preceq B^1\stimes x(k) & \,
\forall k\in\{0,\ldots,K-1\}, \\
        E_\otimes \otimes x(k) \preceq & x(k+1) & & \,
\forall k\in\{0,\ldots,K-1\}
	\end{array}
\end{equation*}
(the third inequality comes from the non-decreasingness of the dater function, i.e., $(\forall k\in\No)\ x(k) = E_\otimes \otimes x(k) \preceq x(k+1)$).
By defining $\tilde{x}_K = [x(0)^\intercal\ x(1)^\intercal\ \dots \ x(K)^\intercal]^\intercal$ and using Proposition~\ref{pr:inequalities}, the system can be rewritten as
\begin{equation}\label{eq:WC}
    M_K \otimes \tilde{x}_K \preceq \tilde{x}_K,
\end{equation}
where
\[
    M_K = 
    \begin{bmatrix}
        C & P & \pazocal{E} & \pazocal{E} & \ldots & \pazocal{E}\\
        I & C & P & \pazocal{E} & \ldots & \pazocal{E}\\
        \pazocal{E} & I & C & P & \ldots & \pazocal{E}\\
        \pazocal{E} & \pazocal{E} & I & C & \ldots & \pazocal{E}\\
        \vdots & \vdots & \vdots & \vdots & \ddots & \vdots \\
        \pazocal{E} & \pazocal{E} & \pazocal{E} & \pazocal{E} & \ldots & C
    \end{bmatrix} \in \Rmax^{n(K+1)\times n(K+1)},
\]
$P = B^{1\sharp}$, $I = A^1\oplus E_\otimes$, $C = A^0\oplus B^{0\sharp}$,\footnote{Matrices $P$, $I$, and $C$ were initially introduced in~\cite{zorzenon2021periodic}, where their name stood respectively for proportional, inverse, and constant matrix. 
Although the original name choice is not meaningful for the present paper, we decided to maintain it here for continuity's sake.} and $\pazocal{E}_{ij} = -\infty$ for all $i,j\in\{1,\ldots,n\}$.
Therefore, a P-TEG is weakly consistent if and only if, for all $K\in\No$,~\eqref{eq:WC} admits a solution $\tilde{x}_K\in\R^{n(K+1)}$ and, if a P-TEG is not weakly consistent, the first $K$ for which the inequality does not admit solutions corresponds to the maximum number of firings before the first token death.

To exploit the results of~\cite{hoeftig1995minimum} for the study of WC we now take, loosely speaking, the limit of the precedence graph $\graph(M_K)$ for $K\rightarrow \infty$.
A rigorous definition for this type of graphs is given as follows.

\begin{defn}[Periodic graph]\label{def:periodic_graph}
    The periodic graph associated with matrices $P,I,C\in\R^{n\times n}$ is the infinite weighted directed graph $\graph(P,I,C) = (\nodes_\infty,\arcs_\infty)$ where $\nodes_\infty = \{1,\ldots,n\} \times \Z$ is the infinite set of nodes, and $\arcs \subseteq \nodes_\infty\times \nodes_\infty$ is the infinite set of arcs, defined such that there is an arc of the form $((j,z),(i,z-1))$ and weight $P_{ij}$ iff $P_{ij}\neq -\infty$, there is an arc of the form $((j,z),(i,z+1))$ and weight $I_{ij}$ iff $I_{ij}\neq -\infty$, and there is an arc of the form $((j,z),(i,z))$ and weight $C_{ij}$ iff $C_{ij}\neq -\infty$.
\end{defn}

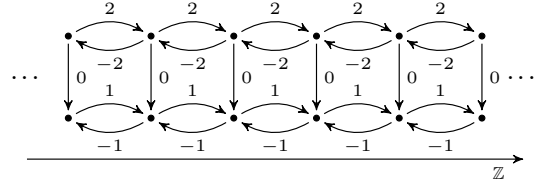
\begin{figure}
	\centering
	\resizebox{.85\linewidth}{!}{
		\begin{tikzpicture}[node distance=2cm and 2cm,>=stealth',bend angle=45,double distance=.5mm,arc/.style={->,>=stealth'},place/.style={circle,thick,minimum size=8mm,draw}]

\tiny

\foreach \z in {1,2,3,4,5,6}
{
\node (ntop\z) at (\z,0) {};
\node (nbot\z) at (\z,-1) {};
\filldraw (ntop\z) circle (1pt);
\filldraw (nbot\z) circle (1pt);
\draw [arc] (ntop\z) to node[auto] {$0$} (nbot\z);
}
\draw [arc] (ntop1) to [bend left=30] node[auto] {$2$} (ntop2);
\draw [arc] (ntop2) to [bend left=30] node[auto] {$2$} (ntop3);
\draw [arc] (ntop3) to [bend left=30] node[auto] {$2$} (ntop4);
\draw [arc] (ntop4) to [bend left=30] node[auto] {$2$} (ntop5);
\draw [arc] (ntop5) to [bend left=30] node[auto] {$2$} (ntop6);
\draw [arc] (ntop2) to [bend left=30] node[auto] {$-2$} (ntop1);
\draw [arc] (ntop3) to [bend left=30] node[auto] {$-2$} (ntop2);
\draw [arc] (ntop4) to [bend left=30] node[auto] {$-2$} (ntop3);
\draw [arc] (ntop5) to [bend left=30] node[auto] {$-2$} (ntop4);
\draw [arc] (ntop6) to [bend left=30] node[auto] {$-2$} (ntop5);

\draw [arc] (nbot1) to [bend left=30] node[auto] {$1$} (nbot2);
\draw [arc] (nbot2) to [bend left=30] node[auto] {$1$} (nbot3);
\draw [arc] (nbot3) to [bend left=30] node[auto] {$1$} (nbot4);
\draw [arc] (nbot4) to [bend left=30] node[auto] {$1$} (nbot5);
\draw [arc] (nbot5) to [bend left=30] node[auto] {$1$} (nbot6);
\draw [arc] (nbot2) to [bend left=30] node[auto] {$-1$} (nbot1);
\draw [arc] (nbot3) to [bend left=30] node[auto] {$-1$} (nbot2);
\draw [arc] (nbot4) to [bend left=30] node[auto] {$-1$} (nbot3);
\draw [arc] (nbot5) to [bend left=30] node[auto] {$-1$} (nbot4);
\draw [arc] (nbot6) to [bend left=30] node[auto] {$-1$} (nbot5);

\node (dots1) at (0.5,-.5) {\textbf{\dots}};
\node (dots2) at (6.5,-.5) {\textbf{\dots}};

\draw [arc] (.5,-1.5) to node[pos=.95,below] {$\Z$} (6.5,-1.5);
\end{tikzpicture}
	}
	\caption{Example of periodic graph $\graph(P,I,C)$.}
	\label{fig:periodic_graph}
\end{figure}

For example, in case $\wZ=\wC$ in Subsection~\ref{su:examples}, we obtain
\[
  P = B^{1\sharp}_\wC = 
  \begin{bmatrix}
    -2 & -\infty\\
    -\infty & -1
  \end{bmatrix},\ I = A^1_\wC \oplus E_\otimes = 
  \begin{bmatrix}
    2 & -\infty\\
    -\infty & 1
  \end{bmatrix},
\] 
\[
  C = A^0_\wC \oplus B^0_\wC = 
  \begin{bmatrix}
    -\infty & -\infty\\
    0 & -\infty
  \end{bmatrix},
\]
and the associated periodic graph is shown in Figure~\ref{fig:periodic_graph}.

The definitions of paths and circuits can naturally be extended from precedence graphs to periodic graphs.
The following lemma relates WC and the circuits of $\graph(P,I,C)$.

\begin{lem}\label{le:WC_and_periodic_graphs}
    A P-TEG is weakly consistent if and only if the periodic graph $\graph(P,I,C)$ does not contain circuits with positive weight.
\end{lem}
\begin{pf}
    First of all, observe that there are no "pathological" periodic graphs where the only circuits of positive weight have infinite length. 
    Suppose that a sequence of circuits with finite length $\rho_k$ from a certain node in $\graph(P,I,C)$ exists such that $\lim_{k\rightarrow \infty} w_{\rho_k} > 0$; then there must also be a circuit $\rho_k$ with positive weight and finite length.
    Indeed, if $(\forall k\in\No)\ w_{\rho_k} \in (-\infty,0]$, we would have a contradiction, as the limit of a sequence of values in a right-closed interval can not be greater than the interval's upper bound.

    The rest of the proof is straightforward: there is a circuit with positive weight (and finite length) in $\graph(P,I,C)$ $\Leftrightarrow$ the same circuit with positive weight appears in $\graph(M_K)$ for $K$ large enough $\Leftrightarrow$ there is $K\in\No$ such that~\eqref{eq:WC} does not admit any solution $\tilde{x}_K\in \R^{n(K+1)}$ (from Proposition~\ref{pr:inequalities_and_graphs}) $\Leftrightarrow$ the considered P-TEG is not weakly consistent.
    \qed
  \end{pf}

\begin{thm}
  WC can be verified in polynomial time.
\end{thm}
\begin{pf}
    From Lemma~\ref{le:WC_and_periodic_graphs}, verifying WC is equivalent to verifying that there are no circuits with positive weight in periodic graph $\graph(P,I,C)$.
    In~\cite[Theorem 4.8]{hoeftig1995minimum}, the authors show that, given $u\in\{1,\ldots,n\}$, $z\in\Z$, there exists a polynomial-time algorithm (with respect to $n$) that decides if a circuit from a node $(u,z)$ of infinite weight exists in $\graph(P,I,C)$.\footnote{In the notation from~\cite{hoeftig1995minimum}, to apply Theorem 4.8 to our problem we need to take $d=1$, $t_{\textup{max}}=1$, $v = u$, $m = 0$.
    Moreover, observe that Theorem 4.8 focuses on circuits with weight $-\infty$; it is however easy to modify the theorem so that its result can be applied for the analysis of circuits with weight $\infty$.
    }
    Since such a circuit exists iff there is a circuit with positive weight from $(u,0)$, it is sufficient to apply the algorithm for all nodes $(u,0)$ with $u\in\{1,\ldots,n\}$ to verify WC.
    \qed
\end{pf}
  
Before we give formulation to the algorithm, we introduce more notation for periodic graphs.
For a node $(i,z)\in\nodes_\infty$ we call $i$ and $z$ its row and column, respectively, and for a path $\rho=((i,z),\ldots,(j,\ell))$, we denote by $t_\rho=\ell - z\in\Z$ the difference between columns of the last and first node of $\rho$. 

The algorithm from~\cite{hoeftig1995minimum}, which is adapted for verifying WC in Algorithm~\ref{alg1}, can be divided into two parts.
In the first part, we start by computing matrices $P,I,C$ from the characteristic matrices $A^0,A^1,B^0,B^1$ of the P-TEG under study. 
Then, for every row $i\in\{1,\ldots,n\}$ of $\graph(P,I,C)$ we construct the set $S_i$ of all pairs $(t,w)$ such that $w$ is maximum weight of all paths of the form $((i,0),\ldots,(i,t))$ (i.e., paths starting and ending in the same row $i$) with length at most $n$; formally, $S_{i}=\{(t,w) \mid w= \text{max}_{\rho \in Q_{i,t}} w_\rho\}$, where $Q_{i,t}= \{\rho \mid \rho=((i,0),\ldots, (i,t)) \wedge |\rho|\leq n\}$.
Since we consider only paths of length at most $n$ in computation of $S_i$, then $t\in\{-n,\ldots,n\}$, and hence there are at most $2n+1$ different pairs in every set $S_i$.
Next, we compute the matrix $R\in\Rmax^{n\times n}$, whose entry $R_{ij}=R_{ji}$ is equal to the maximum weight of all circuits with length at most $n^2$ that pass through both the $i$th and the $j$th row of $\graph(P,I,C)$; formally, $R_{ij}=w_\tau$ if there is circuit $\tau = ((u,0),\ldots,(u,0))$ in $\graph(P,I,C)$ for some $u\in \{1,\ldots, n\}$ such that $|\tau|\leq n^2$ and $\tau$ contains both $(i,z)$ and $(j,\ell)$ for some $z,\ell\in\Z$, and $R_{ij}=-\infty$ otherwise.

In the second part, for each entry $R_{ij}\neq-\infty$ and each pairs $(t,w)\in S_i$ and $(t',w')\in S_j$, the algorithm checks whether there exists a solution $(y,y')\in\No\times \No$ to the following Diophantine system of a linear equation and a linear inequality:
\begin{equation}\label{eq:diophantine}
  \left\{
	\begin{array}{llll}
    yt &+& y't' &= 0\\
		yw &+& y'w' &> 0
	\end{array}
  \right. .
\end{equation}
The existence of a solution $(y,y')$ is a necessary and sufficient condition for the existence of a circuit with positive weight in $\graph(P,I,C)$.
A circuit with positive weight can then be constructed by concatenating $\tilde{y}\in\No$ times a path $\rho\in S_i$ with $t_\rho=t$ and $w_\rho=w$ (opportunely translated, i.e., if $\rho=((i,0),\ldots,(i,t))$, then the concatenation will generate $((i,0),\ldots,(i,t),\ldots,(i,2t),\ldots,(i,\tilde{y}t))$), $\tilde{y}'\in\No$ times a path $\rho'\in S_j$ with $t_{\rho'}=t'$ and $w_{\rho'}=w'$, and connecting the two resulting paths to form a circuit with total weight $\tilde{y}w+\tilde{y}'w'+R_{ij}$, where $(\tilde{y},\tilde{y}')$ is a solution to
\begin{equation}\label{eq:diophantine1}
  \left\{
	\begin{array}{llll}
    \tilde{y}t &+& \tilde{y}'t' &= 0\\
		\tilde{y}w &+& \tilde{y}'w' + R_{ij} &> 0
	\end{array}
  \right. .
\end{equation}

\IncMargin{1.5em}
\begin{algorithm2e}
  \small
  \DontPrintSemicolon
  \caption{Verification of weak consistency}
  \label{alg1}
  \SetKwInOut{Input}{Input}
  \SetKwInOut{Output}{Output}
  \Indmm
  \Input{$A^0,A^1,B^0,B^1\in\Rmax^{n\times n}$}
  
  \Output{{\tt true} iff the P-TEG characterized by $A^0,A^1,B^0,B^1$ is weakly consistent}
  \Indpp
  \BlankLine
  $P = B^{1\sharp},\ I = A^1\oplus E_\otimes,\ C = A^0\oplus B^{0\sharp}$\;
  Compute the set of pairs $S_i$ for every $i\in\{1,\ldots,n\}$\;
  Compute the matrix $R$\; 
  
  \For{$i,j=1$ to $n$}{
    \If{$R_{ij}\neq -\infty$}{
      \For{every $(t,w)\in S_i,\ (t',w')\in S_j$}{
        \If{exist $y,y'\in \mathbb{N}_0$ such that $yt + y't' = 0\, \wedge\, yw + y'w' > 0$}{
          \Return {\tt false}
        }
      }
    }
  }
  \Return {\tt true}
\end{algorithm2e}
\DecMargin{1.5em}

  
  

\citeauthor*{hoeftig1995minimum} show that, using breath-first-search, sets $S_i$ for all $i\in\{1,\ldots n\}$ and matrix $R$ can be obtained in time $\pazocal{O}(n^7)$ and $\pazocal{O}(n^{10})$, respectively.
(Alternatively, it is possible to skip the computation of $S_i$ and $R_{ij}$ by obtaining $M_{2n}^*$ and $M_{2\floor{\frac{n^2}{2}}}^*$ in time $\pazocal{O}(n^6)$ and $\pazocal{O}(n^9)$, from which equivalent information can be extracted.)
Due to its simple form, the Diophantine system~\eqref{eq:diophantine} can be solved analytically in constant time.
Therefore, the total time complexity of the algorithm is $\pazocal{O}(n^{10})$ (or, alternatively, $\pazocal{O}(n^9)$).

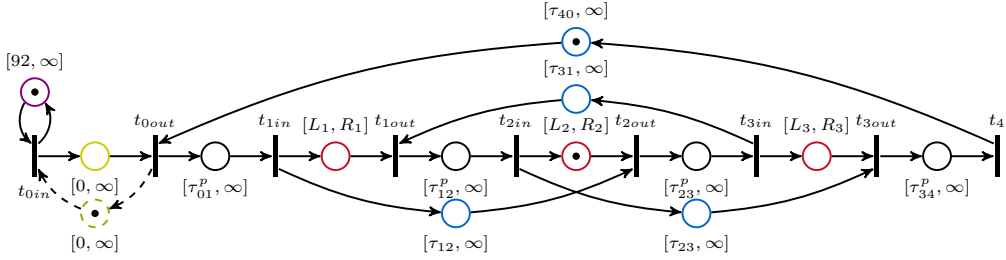
\begin{figure*}[ht]
	\centering
		\centering
		\resizebox{.75\textwidth}{!}{
		\begin{tikzpicture}[node distance=.5cm and .6cm,>=stealth',bend angle=30,thick]
\scriptsize
\node[transitionV,label=above:{$t_{0out}$}] (t0) {};
\node[place,right=of t0,label=below:{$[\tau_{01}^p,\infty]$}] (p01) {};
\node[transitionV,right=of p01,label=above:{$t_{1in}$}] (t1in) {};
\node[place,myred,right=of t1in,label=above:{$[L_1,R_1]$}] (p1) {};
\node[transitionV,right=of p1,label=above:{$t_{1out}$}] (t1out) {};
\node[place,right=of t1out,label=below:{$[\tau_{12}^p,\infty]$}] (p12) {};
\node[transitionV,right=of p12,label=above:{$t_{2in}$}] (t2in) {};
\node[place,myred,tokens=1,right=of t2in,label=above:{$[L_2,R_2]$}] (p2) {};
\node[transitionV,right=of p2,label=above:{$t_{2out}$}] (t2out) {};
\node[place,right=of t2out,label=below:{$[\tau_{23}^p,\infty]$}] (p23) {};
\node[transitionV,right=of p23,label=above:{$t_{3in}$}] (t3in) {};
\node[place,myred,right=of t3in,label=above:{$[L_3,R_3]$}] (p3) {};
\node[transitionV,right=of p3,label=above:{$t_{3out}$}] (t3out) {};
\node[place,right=of t3out,label=below:{$[\tau_{34}^p,\infty]$}] (p34) {};
\node[transitionV,right=of p34,label=above:{$t_4$}] (t4) {};

\node[place,myblue,above=.4cm of p2,label=above:{$[\tau_{31},\infty]$}] (p31r) {};
\node[place,myblue,tokens=1,above=.4cm of p31r,label=above:{$[\tau_{40},\infty]$}] (p40r) {};
\node[place,myblue,below=.4cm of p12,label=below:{$[\tau_{12},\infty]$}] (p12r) {};
\node[place,myblue,below=.4cm of p23,label=below:{$[\tau_{23},\infty]$}] (p23r) {};

\node[place,myyellow,left=of t0,label=below:{$[0,\infty]$}] (p0) {};
\node[transitionV,left=of p0,label=below:{$t_{0in}$}] (t0in) {};
\node[place,mypurple,tokens=1,above=.4cm of t0in,label=above:{$[92,\infty]$}] (p0a) {};

\node[place,dashed,tokens=1,mygreen,below=.4cm of p0,label=below:{$[0,\infty]$}] (pcap) {};

\draw (t0) edge[->] (p01);
\draw (p01) edge[->] (t1in);
\draw (t1in) edge[->] (p1);
\draw (p1) edge[->] (t1out);
\draw (t1out) edge[->] (p12);
\draw (p12) edge[->] (t2in);
\draw (t2in) edge[->] (p2);
\draw (p2) edge[->] (t2out);
\draw (t2out) edge[->] (p23);
\draw (p23) edge[->] (t3in);
\draw (t3in) edge[->] (p3);
\draw (p3) edge[->] (t3out);
\draw (t3out) edge[->] (p34);
\draw (p34) edge[->] (t4);
\draw (t0in) edge[->] (p0);
\draw (p0) edge[->] (t0);

\draw (t3in.180-75) edge[bend right=10,->] (p31r);
\draw (p31r) edge[bend right=10,->] (t1out.75);
\draw (t4.180-75) edge[bend right=10,->] (p40r);
\draw (p40r) edge[bend right=10,->] (t0.75);
\draw (t1in.-75) edge[bend right=10,->] (p12r);
\draw (p12r) edge[bend right=10,->] (t2out.180+75);
\draw (t2in.-75) edge[bend right=10,->] (p23r);
\draw (p23r) edge[bend right=10,->] (t3out.180+75);

\draw (t0in.75) edge[bend right=40,->] (p0a);
\draw (p0a) edge[bend right=40,->] (t0in.180-75);

\draw [dashed] (t0.180+75) edge[bend left=20,->] (pcap);
\draw [dashed] (pcap) edge[bend left=20,->] (t0in.-75);

\end{tikzpicture}
		}
	\caption{P-TEG modeling the electroplating line.
	A token in a place colored \textcolor{myred}{\textbf{red}}, \textcolor{black}{\textbf{black}}, \textcolor{myblue}{\textbf{blue}}, and \textcolor{myyellow}{\textbf{yellow}} represents a part being processed in a tank, the hoist moving with and without carrying a part, and a part waiting in the input depot, respectively.
    The time window attached to the \textcolor{mypurple}{\textbf{purple}} place limits the maximum entrance rate of parts.
    The dashed \textcolor{mygreen}{\textbf{green}} place and arcs model the capacity of the depot.}
	\label{fig:P-TEG_electroplating_line}
\end{figure*}

If the P-TEG under analysis is not weakly consistent, by finding the minimal solution $(\tilde{y}_m,\tilde{y}'_m)$ to~\eqref{eq:diophantine1} (again, such solution is computable in time $\pazocal{O}(1)$) it is possible to obtain an upper bound for the length of the longest consistent finite trajectory.
Since $\tilde{y}_m$, $\tilde{y}_m'$ indicate the number of concatenations of paths $\rho,\rho'$ with $|\rho|,|\rho'|\leq n$ and $t_\rho,t_{\rho'}\in\{-n,\ldots,n\}$, and a circuit with positive weight can be obtained by joining the resulting paths with at most $n^2$ arcs, this circuit certainly belongs to $\graph(M_{\hat{K}})$, where 
\begin{equation}\label{eq:horizon}
    \hat{K}=\tilde{y}_m|t|+2n+2\floor{\frac{n^2}{2}}+1.
\end{equation}
Thus, the first token death occurs for $k\leq \hat{K}$.
From this reasoning, it is trivial to derive a pseudo-polynomial time algorithm that finds the exact $k$ corresponding to the first token death: it is sufficient to implement a binary search (see, e.g., \cite{knuth1998art}) to analyze $\graph(M_k)$, $k\in\{0,\ldots,\hat{K}\}$, until the minimum $k$ for which $\graph(M_k)\notin\nonegset$ is found.
This algorithm terminates in $\pazocal{O}(n^3\hat{K}^3\log\hat{K})$ operations in the worst case.


\section{Practically motivated example}\label{se:5}

To show the usefulness of WC in applications, we present an example of a single-hoist electroplating line with limited input rate.
Consider a system consisting of 3 processing tanks $T_1,T_2,T_3$ and a hoist of capacity one.
Each part entering the system is first deposited in an input depot; the rate at which parts can enter the depot is limited to a maximum of one piece every 92 time units.
After that, parts are grabbed by the hoist, whose task is to transport them from the input depot (denoted $T_0$) to tank $T_1$, then from $T_1$ to $T_2$, from $T_2$ to $T_3$, and finally from $T_3$ to an output depot (denoted $T_4$).
To obtain the desired properties, the processing time of each part in tank $T_i$ must be within the interval $\iota_i = [L_i,R_i]\subset \R_{\geq 0}$, for $i\in\{1,2,3\}$.
The time it takes the hoist to move from $T_i$ to $T_{i+1}$ while carrying a part is denoted $\tau^p_{i,i+1}$, for $i \in\{0,1,2,3\}$.

With the aim to increase its throughput, the system is able to process more than one part at a time: while a part is being processed in a tank $T_i$, another one may reside in tank $T_j$, with $i\neq j$.
To synchronize the treatments, the hoist shall thus move between two tanks $T_i$ and $T_j$ without carrying a part, so that it can grab the part residing in $T_j$; this movement takes $\tau_{ij}$ time units.
We suppose the hoist is programmed to follow a cyclic sequence of operations: let "$T_i \xrightarrow{p} T_{i+1}$" indicate the transportation of a part from $T_i$ to $T_{i+1}$, and "$\rightarrow T_{i}$" denote the movement of the hoist from its previous location to $T_i$.
Then, the cyclic hoist operation is described by repetitions of the following sequence of tasks: 
\[
	\rightarrow T_0 \xrightarrow{p} T_1 \rightarrow T_2 \xrightarrow{p} T_3 \rightarrow T_1 \xrightarrow{p} T_2 \rightarrow T_3 \xrightarrow{p} T_4.
\]
Observe that at the beginning and the end of each repetition of the sequence, one part is left in tank $T_2$.

The following numerical parameters are used: $\tau_{ij} = |i-j|,\ \tau_{ij}^p = \tau_{ij}+1,\ \iota_1 = [20,30],\ \iota_2 = [25,35],\ \iota_3 = [20,30]$.
The P-TEG of Figure~\ref{fig:P-TEG_electroplating_line} models the electroplating line.
We consider two cases: (i) the depot has infinite capacity (the dashed arcs and place in Figure~\ref{fig:P-TEG_electroplating_line} are not considered); (ii) the depot has capacity one (the dashed arcs and place are considered).
The execution of Algorithm~\ref{alg1} shows that only in case (i) the P-TEG is weakly consistent.
In case (ii) the algorithm finds that, for $i=1$, $j=3$, $R_{ij}=-73$, $(t,w)=(1,92)\in S_i$, $(t',w')=(-9,-819)\in S_j$, a solution $(y,y')$ to~\eqref{eq:diophantine} exists, and that the minimal solution to~\eqref{eq:diophantine1} is $(\tilde{y}_m,\tilde{y}_m)=(81,9)$; from~\eqref{eq:horizon}, this implies that for a certain $k\leq \hat{K} = 180$ the first token death occurs.
In fact, a more careful analysis reveals that the first token death occurs at the 119th firing, as $\graph(M_{118})\notin\nonegset$; therefore, the electroplating line can process at most 118 parts before a part is left for too long in one of the tanks for the first time, forcing the system to restart.
Instead, in case (i), no solution to~\eqref{eq:diophantine} is found; hence, although the system is not suitable to be used in continuous production (as the P-TEG is not boundedly consistent), the electroplating line can be set up to process the desired amount of parts for any customer's demand.
A consistent finite trajectory that can be used to process $K\in\N$ parts can be found as 
\[
	\tilde{x}_K=[x(0)^\intercal\ x(1)^\intercal\ \dots \ x(K)^\intercal]^\intercal=M_K^*\otimes u,
\]
where $u$ is any vector from $\R^{n(K+1)}$ (see Proposition~\ref{pr:inequalities_and_graphs}). 

Despite its significant computational complexity, a Matlab implementation of Algorithm~\ref{alg1} on a PC with an Intel i7 processor at 2.20Ghz verifies WC on these examples of P-TEGs in only 0.15 seconds.
Finding the $k$ corresponding to the first token death in case (ii) takes 4.5 seconds using the binary search algorithm.
\section{Final remarks}\label{se:conclusions}

We have proposed an algorithm of strongly polynomial time complexity for verifying WC, and an algorithm of pseudo-polynomial time complexity for finding the length of the longest consistent finite trajectory for a P-TEG.
Two interesting open questions are left for future work:
\begin{enumerate}
    \item is there an algorithm that checks consistency?
    \item is there a polynomial-time algorithm that finds the maximum number of firings before the first token death?
\end{enumerate}

\bibliography{references}

\end{document}